\documentclass[twocolumn,prb,aps,superscriptaddress,showpacs,amsmath,amssymb]{revtex4}
\usepackage{amsfonts}
\usepackage{bbm}

\usepackage{graphicx}
\usepackage{dcolumn}
\usepackage{bm}

\begin{document}

\title{Enhanced spin-polarized transport through DNA double helix by gate voltage}
\author{Ai-Min Guo}
\affiliation{Institute of Physics, Chinese Academy of Sciences, Beijing 100190, China}
\author{Qing-feng Sun}
\affiliation{Institute of Physics, Chinese Academy of Sciences, Beijing 100190, China}

\date{\today}

\begin{abstract}
We report on a way to manipulate the spin transport through double-stranded DNA contacted by normal-metal electrodes. On the basis of an effective model Hamiltonian, the conductance and the spin polarization are calculated in the presence of a gate voltage by using the Landauer-B\"{u}ttiker formula. Our results indicate that the spin polarization presents strong dependence on the magnitude as well as the direction of the gate voltage. The spin polarization can be significantly enhanced by tuning the gate voltage and shows oscillating behavior with increasing the DNA length.
\end{abstract}


\maketitle

\section{\label{sec1}Introduction}

The field of spintronics, which aims at using the electron spin to store and process information, has triggered extensive interest during the last two decades.\cite{zi} Since the discovery of the giant magnetoresistance in 1988,\cite{bmn} much progress has been achieved on the spin transport through solid-state systems and a set of spintronic devices were proposed based on organic materials. Magnetic tunnel junctions were fabricated from organic semiconductor and spin injection across metal-organic interface was demonstrated.\cite{bc} A supramolecular spin-valve device was presented by coupling single molecule magnets to a single-walled carbon nanotube quantum dot.\cite{um} The spin transport properties of the DNA molecule were investigated theoretically by connecting to ferromagnetic electrodes.\cite{zm,wxf} The spin effects of all these systems arise from magnetic materials and from heavy atoms with large spin-orbit interactions, and are not determined by the organic molecules themselves.

Recently, an efficient spin filter was reported by depositing self-assembled monolayers of double-stranded DNA (dsDNA) on gold substrate\cite{gb} or by sandwiching single dsDNA between two electrodes.\cite{xz} The electrons are highly polarized after transmitting through the dsDNA with spin polarization up to 60\% at room temperature. Moreover, the spin filtration efficiency increases with the DNA length, implying that the spin effects are dominated by the dsDNA and do not depend on the interface between the dsDNA and the gold surface. These results are surprising since the DNA molecule is nonmagnetic and has weak spin-orbit coupling (SOC) that could not support such high spin polarization. Until now, several theoretical works have studied the spin-polarized transport through single-stranded DNA (ssDNA) based on the helical chain-induced Rashba SOC.\cite{ys,gr} Very recently, we proposed a model Hamiltonian to explain the experiment by combining the SOC, the dephasing, and the double helix structure of the DNA molecule.\cite{gam} The results indicated that the spin polarization is significant for the dsDNA even in the case of small SOC and increases with its length, while no spin polarization occurs in the ssDNA. These are in good agreement with the experimental results.\cite{gb,xz}

The DNA molecule is a promising candidate for molecular electronics (see Refs.~\onlinecite{erg,gjc} for review), due to its unique structural and self-assembling properties. As compared with conventional semiconductors and metals, the DNA molecule preserves long spin relaxation time which makes it attractive for building spintronic devices. Meanwhile, it was reported that the dsDNA could be a field-effect transistor in the presence of a gate electrode.\cite{ykh,mav} Consequently, one may ask the following questions: (1) will the gate voltage affect the spin transport of the dsDNA? (2) can the gate voltage be used to control its spin transport? Besides, the components in each integrated circuit have different electric potentials. It is thus important to illustrate how these potentials will influence the spin transport of the dsDNA-based devices.

\begin{figure}
\includegraphics[scale=0.8]{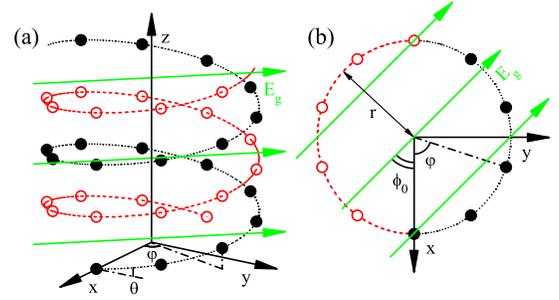}
\caption{\label{fig:zero}(color online). (a) Schematic view of right-handed dsDNA with helix angle $\theta$ and cylindrical coordinate $\varphi$ under external electric field $E_g$. The left-handed dsDNA can be derived by replacing $\theta$ with $\pi- \theta$ and $\varphi$ with $ -\varphi$. (b) Projection of bottom five base-pairs and the electric field into $x$-$y$ plane. Here, $r$ and $\phi_0$ are radius of the dsDNA and angle from the electric field to the bottommost (first) base-pair (negative direction of $x$ axis), respectively. The spin-selectivity of the dsDNA originates from the combination of the SOC, the dephasing, and its double helix structure.\cite{gam} Moreover, the spin filtration efficiency could be considerably enhanced by adjusting the magnitude and the direction of $E_g$ (see text).}
\end{figure}

In this paper, we investigate the spin-selective tunneling of electrons through the dsDNA contacted by nonmagnetic electrodes in the presence of an external electric field, which is perpendicular to the helix axis of the dsDNA, as illustrated in Fig.~\ref{fig:zero}. On the basis of an effective model Hamiltonian, the conductance and the spin polarization are calculated by using the Landauer-B\"{u}ttiker formula. We find that the spin polarization strongly depends on the magnitude as well as the direction of the gate field. The dsDNA could be a very efficient spin filter and the spin polarization is considerably large under the gate voltage. Furthermore, the spin polarization exceeds 70\% for long dsDNA by properly tuning the gate voltage.

The rest of the paper is organized as follows. In Sec.~\ref{sec2}, the model is presented. In Sec.~\ref{sec3}, the spin polarization and the conductance are shown in the presence of the gate voltage. Finally, the results are summarized in Sec.~\ref{sec4}.

\section{\label{sec2}Model}

The spin transport through the dsDNA can be simulated by the
Hamiltonian:\cite{gam}
\begin{eqnarray}
\begin{aligned}
{\cal H}= & \sum_j (\sum_n \varepsilon_ {jn} c_{jn}^\dag c_{jn}+ \sum _{n=1}^{N-1} t_{j}c_{jn}^\dag c_{jn+1}  +\mathrm{H.c.}) \\& +\sum_n( \lambda c_{1n}^\dag c_{2n} +\mathrm{H.c.}) \\& + \sum_{j n}\{ i t_{\rm so} c_{jn}^\dag [ \sigma _n ^ {(j)}+ \sigma_{n+1}^{(j)}] c_{jn+1}+ \mathrm{H.c.} \} \\& +\sum_{jnk} ( \varepsilon_{jnk}b_{jnk}^\dag b_{jnk}  + t_d b_{jnk}^ \dag c_{jn} +\mathrm{H.c.}) \\ & + \sum_{jk} ( t_L a_{L k}^\dag c_{j 1} +t_R a_{R k} ^ \dag c_{j N} + \mathrm {H.c.} ) \\ & + \sum_{k, \beta=L,R} \varepsilon_{\beta k} a_{\beta k}^\dag a_{\beta k}. \label{eq:one}
\end{aligned}
\end{eqnarray}
Here the first two terms are the Hamiltonian of usual two-leg ladder model including the spin degree of freedom. $c_{jn} ^\dag= (c_{jn \uparrow} ^\dag, c_{ jn \downarrow } ^\dag)$ is the creation operator of the spinor, with $j=1,2$ labeling a strand and $n\in [1,N]$ denoting a base-pair of the dsDNA. $\varepsilon_ {jn} $ is the on-site energy, $t_j$ is the intrachain hopping integral, and $\lambda $ is the interchain hybridization interaction. The third term is the SOC Hamiltonian, arising from the double helix shape of the electrostatic potential of the dsDNA.\cite{gam} $t_{\rm so}$ is the SOC and $\sigma_{n+1} ^{(j ) } =\sigma_z \cos\theta- (-1)^j [\sigma_ x \sin \varphi - \sigma_ y \cos \varphi] \sin \theta$, with $\sigma_ {x,y,z} $ the Pauli matrices, $\theta$ the helix angle, $\varphi= n\Delta \varphi$ the cylindrical coordinate, and $\Delta\varphi$ the twist angle. The fourth one denotes the B\"{u}ttiker's virtual electrodes,\cite{zhy,xy} which are introduced to simulate the phase-breaking processes by attaching each base to a virtual electrode,\cite{bya1,bya2} because of the inelastic scattering from the phonons and other inelastic collisions with the counterions. The last two terms describe the coupling between the real nonmagnetic electrodes and the dsDNA, and the real electrodes, respectively.

When the dsDNA is subjected to a perpendicular electric field (Fig.~\ref{fig:zero}), the on-site energy at each base site will be modulated into following form:
\begin{equation}
\varepsilon_ {jn}=\varepsilon_ {jn}^{(0) }- (-1)^j eV_g \cos[ (n-1) \Delta\varphi + \phi_0], \label{eq:two}
\end{equation}
where $\varepsilon_ {jn}^{(0) }$ is the on-site energy of the base at zero electric field and $e$ is the elementary charge. $V_g =E_g r$ is the gate voltage across the dsDNA with $E_g$ the perpendicular electric field and $2r$ the effective distance between the complementary bases. The phase $\phi_0$, being the angle between the electric field and the first base-pair, reflects the orientation of the gate voltage with respect to the dsDNA, as seen in Fig.~\ref{fig:zero}. $\phi_0$ could be changed by rotating the dsDNA with the direction of its helix axis fixed. Besides the gate electrode,\cite{ykh} the gate voltage may also originate from the voltage drop across the left and right real electrodes.\cite{mav} One notices from Eq.~(\ref{eq:two}) that the gate voltage tunes the on-site energies harmonically along each strand and introduces disorder within each pitch of the dsDNA, due to the intrinsic double helix structure of the dsDNA. Such modulation will definitely modify the electronic structure of the DNA molecule and thus influences its transmission ability as well as the spin polarization (see below). Finally, the gate voltage is chosen to be the order of $0.1$ volt, where the external electric field is much smaller than the internal one produced by the nuclei of the dsDNA and will not contribute to the SOC.

The current in the $q$th real or virtual electrode with spin $s= \uparrow,\downarrow$ can be obtained from the Landauer-B\"{u}ttiker formula $I_{q s}=(e^2/h) \sum_{m, s'}T_{qs,ms'} (V_{m}- V_{q})$, where $V_q$ is the voltage in the $q$th electrode and $T_{qs,ms'}$ is the transmission coefficient from the $m$th electrode with spin $s'$ to the $q$th electrode with spin $s$.\cite{gam} By applying a small bias between the real electrodes with $V_L =V_b $ and $V_R=0$, $V_q$ can be derived for the virtual electrodes, since the net current flowing through each of them is zero. Then the conductances for spin-up and spin-down electrons can be calculated $G_{s} =(e^2/h) \sum_{m, s'}T_{Rs,ms'} V_ {m}/ V_b $. The spin polarization is $P_s=(G_ \uparrow- G_ \downarrow )/ (G_ \uparrow+ G_\downarrow)$.

The values of aforementioned parameters are the same as those in Ref.~\onlinecite{gam}, i.e., $\varepsilon_ {1n}^{(0)}= 0$, $\varepsilon_ {2n}^{(0)} = 0.3 $, $t_{1}=0.12$, $t_{2}=-0.1$, and $\lambda=-0.3$, which are determined from first-principles calculations\cite{zhy,sk,hlgd} and the unit is eV. Other parameters are $t_{\rm so}=0.01$, $\theta = 0.66$ rad, and $\Delta\varphi ={\frac \pi 5}$, indicating that there are ten base-pairs within the pitch of the dsDNA. For the real electrodes, the linewidth functions are $\Gamma _ {L/R} =1$; for the virtual ones, the dephasing is small with $\Gamma_d= 4 \times 10^{-4}$ or $4 \times 10^{-3}$,\cite{gam} because the DNA length is shorter than the persistence length\cite{cyf} and the DNA molecule is rigid. When $\Gamma_ d=4\times 10 ^{-3}$, the phase coherence length is $L_ \phi=16$, at which the coherent conductance is equal to the incoherent one.\cite{xy} For short dsDNA of $N<L_ \phi$, the coherent conductance is larger than the incoherent one and the charge transport is determined by quantum mechanism; for the dsDNA of $N>L_ \phi$, the coherent conductance is smaller and the incoherent charge transport mechanism becomes dominant, in accordance with previous results.\cite{bya1,bya2} The effects of the gate voltage on the spin transport through the dsDNA can be observed in both coherent and incoherent charge transport regime (see below), and are generic for different model parameters. In what follows, we mainly focus on right-handed dsDNA except for Fig.~\ref{fig:two}(a), where the spin polarization is shown for both right-handed and left-handed dsDNA.

\section{\label{sec3}Results and Discussions}

Although the gate voltage is employed, no spin polarization could appear also in the ssDNA\cite{gam} and in the dsDNA if any factor of the SOC, the dephasing, and the chirality is absent, regardless of the strength and direction of the gate field. It is well known that the SOC will give rise to spin precession as the charges move.\cite{sqf} For the ssDNA, the charges can transport from one base to another along only single channel. In this case, the Hamiltonian can be transformed into a spin-independent one by using a unitary transformation,\cite{gam} which is equivalent to choosing a space-dependent spin-rotating frame.\cite{sqf} In this spin-rotating frame which follows the spin precession, the spin is invariant. Accordingly, the Hamiltonian does not depend on the spin after the transformation and no spin polarization could be obtained in the ssDNA. Nevertheless, the dsDNA presents a fundamental distinction. There are many channels for the charges to propagate between two bases, and one cannot find a spin-rotating frame to follow the spin precession. As a result, the Hamiltonian cannot be altered into a spin-independent one and the spin polarization will appear.

\begin{figure}
\includegraphics[scale=0.6]{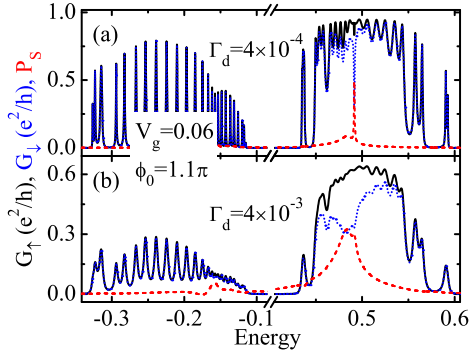}
\caption{\label{fig:one}(color online). Energy-dependent conductances $G_{\uparrow}$ (solid line), $G_{\downarrow}$ (dotted line), and spin polarization $P_s$ (dashed line) in the presence of the gate voltage for (a) $\Gamma_d=4\times10^{-4}$ and for (b) $\Gamma_d=4\times10^{-3}$ with $N=30$.}
\end{figure}

Figures~\ref{fig:one}(a) and~\ref{fig:one}(b) plot the conductances $G_{\uparrow/\downarrow}$ and the spin polarization $P_s$ under the gate voltage with $V_g$$=$$0.06$ and $\phi_0$$=$$1.1\pi$ for two dephasing strength $\Gamma_d$, as a function of the energy $E$. One notes that the energy spectrum consists of HOMO and LUMO bands which are divided by an energy gap, irrespective of $\Gamma_d$. In comparison with the case of $V_g=0$ (see Fig. 2(a) in Ref.~\onlinecite{gam}), it clearly appears that both bands become fragmented in the presence of the gate voltage, because the periodicity of the system extends to ten base-pairs due to the harmonic variation of the on-site energies. Besides, several transmission peaks are observed in both bands and are more distinct in the regime of smaller $\Gamma_d$ [Fig.~\ref{fig:one}(a)], owing to the stronger coherence of the system. The conductances $G_{\uparrow}$ and $G_{\downarrow }$ are declined by increasing $\Gamma_d$, because the inelastic scattering of the electrons becomes stronger for larger $\Gamma_d$.\cite{gam} On the other hand, one notices a peak or bell-shaped configuration in the curve of $P_s$ vs $E$. The width of $P_s$-$E$ is enhanced by increasing $\Gamma_d$, while its height is reduced.

\begin{figure}
\includegraphics[scale=1.0]{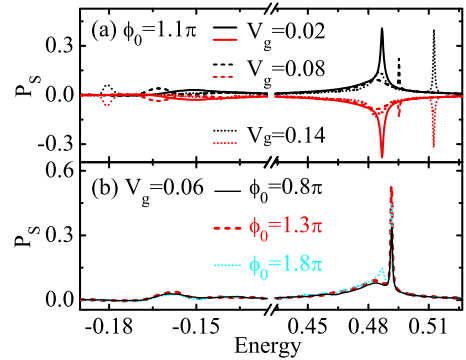}
\caption{\label{fig:two}(color online). $P_s$ vs energy for (a) three values of $V_g$ with $\phi_0=1.1\pi$ and for (b) different $\phi_0$ by fixing $V_g=0.06$ with $\Gamma_d = 4 \times10^ {-4}$ and $N=30$. The black and red curves in Fig.~\ref{fig:two}(a) refer to the right-handed and left-handed dsDNA, respectively.}
\end{figure}

Figure~\ref{fig:two}(a) shows $P_s$ vs $E$ with different values of $V_g$ by fixing $\phi_0 = 1.1\pi$ for the right-handed dsDNA (black curves) and the left-handed one (red curves), which can be obtained by employing the replacement $\theta $$\rightarrow$$\pi - \theta$ and $\varphi$$\rightarrow$$ - \varphi$. By increasing $V_g$, one can see the following features: (1) the peak in the HOMO (LUMO) band is shifted towards lower (higher) energies, since both bands move away from the energy gap; (2) the width and the magnitude of the peak are varied; (3) a new peak will emerge in the LUMO band and locates at the position of the one of $V_g=0$. In addition, the statement that only the sign of the spin polarization will be changed if the chirality of the dsDNA is reversed,\cite{gam} i.e., $P_s(\pi- \theta, -\varphi) =- P_s (\theta,\varphi)$, does not hold in the case of the gate voltage. We consider $P_s$ at $E=0.487$. For the right-handed dsDNA, $P_s$ is $39\%$, $8.0\%$, and $12\%$ for $V_g =0.02$, $0.08$, and $0.14$, respectively; for the left-handed one, $P_s$ is changed to $-37\%$, $-7.6\%$, and $-10\%$, respectively. This is attributed to the broken mirror-symmetry between the right-handed dsDNA system and the left-handed one due to the existence of identical gate field. However, this symmetry will be recovered by properly modulating the direction of the gate field applied on the left-handed dsDNA that the angle between this new field and the first base-pair of the left-handed dsDNA is altered to be $-\phi_ 0$. In this situation, we obtain the relation $P_s(\pi-\theta, -\varphi, -\phi_0)=- P_s ( \theta, \varphi,\phi_0) $. We then investigate the spin polarization by fixing $V_g$, as illustrated in Fig.~\ref{fig:two}(b). A different behavior is observed that the position of the peak remains still, while its magnitude dramatically depends on $\phi_0$. In other words, the spin polarization could be enhanced by adjusting the direction of the gate field with respect to the dsDNA. This arises from the fact that the positions of both HOMO and LUMO bands are unchanged and the details of the energy spectrum are modified due to the rearrangement of the on-site energies. As a result, the conductances will be changed, which leads to the variation of $P_s$.

\begin{figure}
\includegraphics[scale=0.77]{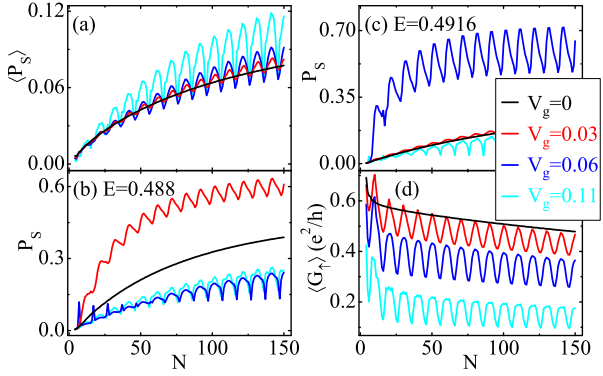}
\caption{\label{fig:three}(color online). Length-dependent (a) $\langle P_s\rangle$, (b) $P_s$ at $E=0.488$, (c) $P_s$ at $E=0.4916$, and (d) $\langle G_{\uparrow}\rangle$ for several values of $V_g$ with $\phi_0 =1.1\pi$ and $\Gamma_d=4\times10^{-4}$.}
\end{figure}

In the following, we calculate the averaged spin polarization $\langle P_s\rangle$, where $\langle P_s \rangle \equiv ( { \langle G_\uparrow \rangle-\langle G_\downarrow \rangle })/ ({ \langle G_\uparrow \rangle + \langle G_\downarrow \rangle})$ with $\langle G_s \rangle$ averaged over the LUMO band. Figure~\ref{fig:three}(a) plots $\langle P_s\rangle$ vs $N$ for several values of $V_g$ with the length up to $N=150$. In comparison with the case of $V_g =0$ that $\langle P_s \rangle$ increases monotonically with $N$,\cite{gam} the dependence of $\langle P_s \rangle$ on $N$ is complicated under the gate voltage. For $V_g \neq 0$, $\langle P_s\rangle$ oscillates between two envelopes, corresponding to the local maxima and minima of $\langle P_s \rangle$ within each pitch of the dsDNA. We find that $\langle P_s \rangle$ usually increases with $N$ if $(N-1) \Delta \varphi+ \phi_0 \in [2m\pi, (2m+1) \pi]$ and decreases with $N$ if $(N-1) \Delta \varphi+ \phi_0 \in [(2m-1)\pi, 2m \pi]$ in every pitch with $m$ the integer, because of the different quantum interference properties in specific length region due to the gating effects. The values of both envelopes and the oscillating amplitude of $\langle P_s \rangle$ increase with $N$. Besides, the oscillating amplitude of $\langle P_s \rangle$ is considerably enhanced by $V_g$. These imply that the spin filtration efficiency can be improved significantly by implementing a perpendicular electric field to long dsDNA.

A similar oscillating behavior can be also observed in the curve of $P_s$ vs $N$ for single Fermi energy, as illustrated in Figs.~\ref{fig:three}(b) and \ref{fig:three}(c). It can be seen that $P_s$ is very sensitive to $V_g$ and $E$. For instance, when $E=0.488$ and $N=92$, $P_s$ is $31\%$, $59\%$, $18\%$, and $21\%$, respectively, by increasing $V_g$ from $0$ to $0.11$; when $N=92$ and $V_g = 0.06$, $P_s$ is increased from $18\%$ at $E=0.488$ to $69\%$ at $E=0.4916$. Figure~\ref{fig:three}(d) shows the averaged conductance $\langle G_{\uparrow} \rangle$ vs $N$. The behavior of $\langle G_{\uparrow} \rangle$ on $N$ has almost the same trend in Fig.~\ref{fig:three}(a), i.e., the specific dependence of the physical quantity on $N$ in different length region. For $V_g \neq 0$, $\langle G_{\uparrow} \rangle$ oscillates between two envelopes corresponding to the local maxima and minima of $\langle G_{ \uparrow} \rangle$ in each pitch of the dsDNA, due to the harmonic modulation of the on-site energies. Both envelopes and $\langle G_{\uparrow} \rangle$ of $V_g=0$ are declined by increasing $N$ and $\langle G_{\uparrow} \rangle$ decreases with $V_g$, since larger $N$ or $V_g$ will strengthen the scattering of the electrons. However, $\langle G_{ \uparrow } \rangle$ remains quite large even for $N=150$ and $V_g = 0.11$, because of the incoherent charge transport mechanism.\cite{bya2} Therefore, the dsDNA could be a better spin filter by modifying the magnitude of the gate voltage.

\begin{figure}
\includegraphics[scale=0.77]{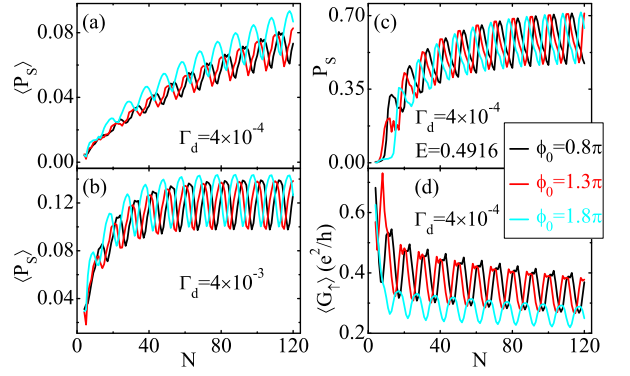}
\caption{\label{fig:four}(color online). Length-dependent (a) and (b) $\langle P_s\rangle$, (c) $P_s$ at $E=0.4916$, and (d) $\langle G_{\uparrow}\rangle$ for several values of $\phi_0$ and $\Gamma_d$ by fixing $V_g =0.06$.}
\end{figure}

We then study the influence of $\phi_0$ on the length-dependent spin
polarization by fixing $V_g$. Figures~\ref{fig:four}(a) and
\ref{fig:four}(b) plot $\langle P_s\rangle$ vs $N$ with three
values of $\phi_0$ and $V_g=0.06$ for $\Gamma_d= 4 \times
10^{-4}$ and $4 \times 10^{-3}$, respectively. Although a similar
behavior is found that $\langle P_s\rangle$ oscillates between
two envelopes, the positions of the local maxima (minima) are
shifted towards smaller $N$ by increasing $\phi_0$, independent of
$\Gamma_d$. For instance, one local maximum for long dsDNA is
decreased from $92$ to $88$ by increasing $\phi_0$ from $0.8\pi$ to
$1.8\pi$. The values of both envelopes increase with $N$ in a wider
range of $N$ in the case of extremely small $\Gamma_d$, and increase
with $N$ at first and are then suppressed by further increasing $N$
for relatively large $\Gamma_d$, because the electron loses its
phase and spin memory faster by increasing $\Gamma_d$.\cite{gam}
Moreover, the magnitude of both envelopes can be enhanced by varying
$\phi_0$ (see the curves of $\phi_0=1.8\pi$), due to the quantum
interference properties. Figure~\ref{fig:four}(c) shows $P_s$
vs $N$ for single Fermi energy, where one notes that $P_s$ is
larger than $70\%$ for long dsDNA [see also Fig.~\ref{fig:three}(c)]
and is almost the same as that observed in the photoelectrons
emitted from strained InGaAs layers.\cite{tm} In addition,
Fig.~\ref{fig:four}(d) shows $\langle G_{\uparrow} \rangle$ vs $N$.
One can see that the conductance remains quite large for $N=120$.
Accordingly, the dsDNA is a very efficient spin filter in the
presence of the gate voltage.

\begin{figure}
\includegraphics[scale=0.6]{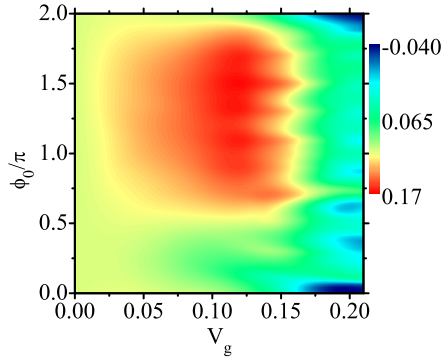}
\caption{\label{fig:five}(color online). Two-dimensional plot of
$\langle P_s \rangle $ vs $V_g$ and $\frac {\phi_0} \pi$
with $\Gamma_d=4 \times10^{-3}$ and $N=100$. It is clear that
$\langle P_s \rangle $ is independent of $\phi_0$ if $V_g =0$ and has the same value for $\phi_0$ and $\phi_0 +2\pi$.}
\end{figure}

Let's further study the spin effects of the dsDNA by varying the magnitude and the direction of the gate field in a wider parameter's range, as plotted in Fig.~\ref{fig:five}. It clearly appears that $\langle P_s\rangle$ increases with $V_g$ at first and is then declined by further increasing $V_g $. $\langle P_s\rangle$ is about $11\%$ at $V_g=0$ and is larger than 16\% by increasing $V_g$ within the range $[0.1,0.13]$. $\langle P_s\rangle$ is very small in the regime of $V_g> 0.2$, where the conductance is also quite small, because of the strong gating effects. On the other hand, the behavior of $\langle P_s \rangle$ on $\phi_0$ is more complex and will have multi-turning points in the curve of $\langle P_s \rangle$-$ \phi_0 $ by fixing $V_g$. For small $V_g$ ($V_g <0.1$), $\langle P_s\rangle$ decreases with $\phi_0$ at first, then increases with $\phi_0$, and is finally decreased by further increasing $\phi_0$; while for large $V_g$ ($V_g>0.13$), $\langle P_s\rangle$ will oscillate with increasing $\phi_ 0$.

\section{\label{sec4}Conclusions}

In summary, we investigate the quantum spin transport through the dsDNA contacted by nonmagnetic electrodes under the gate voltage. This dsDNA-based device could be a very efficient spin filter and the spin filtration efficiency can be improved significantly by modulating the magnitude and the direction of the gate voltage. Our results could be readily checked by further experiments.

\section*{Acknowledgments}

This work was supported by China-973 program and NSF-China under Grants No. 10974236 and 11121063.

\end{document}